\documentclass{aa}
\usepackage{graphicx}
\usepackage{txfonts}
\begin{document}      
   \title{The XMM-$\Omega$ project~: II. Cosmological implications from the high redshift $L-T$ relation of X-ray clusters } 
%
 
       \titlerunning{The XMM-$\Omega$ project: II}  
 
   \author{ 
S.C.~Vauclair 
\inst{1} \and
 A.~Blanchard 
\inst{1} \and
 R.~Sadat 
\inst{1} \and
J.G.~Bartlett
\inst{2,7} \and
J.-P.~Bernard
\inst{3} \and
M.~Boer  
\inst{3} \and 
M.~Giard 
\inst{3} \and 
D.H.~Lumb 
\inst{4} \and 
 P.~Marty 
\inst{5}   \and         
J.~Nevalainen
\inst{6}  
}

   \institute{
Laboratoire d'astrophysique de l'observatoire Midi-Pyr\'en\'ees, CNRS, UMR 5572, UPS,  14, Av. E. Belin, 
31\ 400 Toulouse, France 
\and   
APC, universit\'e Denis Diderot, Paris VII 75\ 005 Paris, France 
\and
Centre d'\'etude spatiale des rayonnements, OMP, UPS, 9, Av. du Colonel Roche, BP4346, 31\ 028 Toulouse, France 
\and
Advanced Concepts and Science Payloads Office, European Space Agency,
ESTEC, 2200AG Noordwijk, Netherlands 
\and
Institut d'astrophysique spatiale,  universit\'e Paris-sud, 91\ 405 Orsay cedex, France
\and
Harvard-Smithsonian Center for Astrophysics, 60 Garden Street, Cambridge, MA02138, USA
\and
Observatoire astronomique de Strasbourg, CDS,  ULP, 11, rue de l'universit\'e
67\ 000 Strasbourg, , France
} 
   
\date{Received \rule{2.0cm}{0.01cm} ; accepted \rule{2.0cm}{0.01cm} }

\abstract{
The evolution with redshift of the temper\-ature-luminosity relation of X-ray galaxy clusters
 is a key ingredient to break degeneracies
in the interpretation of X-ray clusters redshift number counts. 
We therefore take advantage of the recent measurements
of the temperature-luminosity relation of distant clusters
observed with XMM-Newton and Chandra satellites to examine theoretical number counts expected 
for different available X-rays cluster samples, namely the RDCS, EMSS, SHARC, 160deg$^2$ and the MACS at redshift 
greater than 0.3. We derive these counts without any adjustment, using
 models previously normalized  to the local  ($z\sim 0.05$) temperature distribution function 
(TDF) and to the 
high-$z$ ($z \sim 0.33$) TDF. We find that these models having $\Omega_M$ in the range [0.85-1.] predict counts  
in remarkable agreement with the observed counts in the different samples.
 We 
illustrate  that this conclusion is weakly sensitive to the various ingredients of the modeling.
Therefore number counts provide a robust evidence of an evolving population. 
A   realistic flat low density model  ($\Omega_M = 0.3$), 
normalized to the local abundance of clusters is found to overproduce cluster 
abundance at high redshift (above $z \sim 0.5$) by nearly an order of magnitude.
This result is in conflict with the popular concordance model.  The conflict could indicate
a deviation from the expected scaling of the $M-T$ relation with redshift. 
\keywords{Cosmology -- cosmological parameters -- dark matter -- Galaxies: clusters : general}}
 \maketitle

\section{Introduction} 
 
The evolution of the temperature distribution function (TDF) of X-ray clusters is known to be a 
powerful cosmological test of the density parameter of 
the Universe (Oukbir \& Blanchard 
1992). Indeed, the evolution with redshift of the mass function,
 on cluster scales, once normalized to present day,
appears to be a robust  cosmological test, with an exponential sensitivity
to $\Omega_M$ through the gravitational growth rate of perturbations 
(Blanchard \& Bartlett 1998). The high sensitivity of this test has 
allowed its first direct application from a sample  of ten (revised to 9) clusters at redshift 0.3 with measured temperature
(Henry 1997). However, several authors have inferred somewhat 
different
 values from this single sample (Eke et al. 1998; Viana \& Liddle 1999; 
Blanchard et al. 2000, hereafter B00). This might not be so surprising
given the small size of the sample as well as the limited range of redshift.
An alternative approach to track the evolution of the abundance of
clusters is from their redshift 
distribution in  X-ray samples for which
the selection function is known. This procedure, reducing  
 the time telescope investment,
has been applied to the EMSS sample 
(Oukbir \& Blanchard 1997; Reichart et al. 1999) and to the RDCS sample
(Borgani et al. 1999, 2001) but the conclusion on $\Omega_M$ depends
  on the possible evolution 
of the $L-T$ relation (Sadat et al. 1998, hereafter SBO98; 
Borgani et al. 1999, Novicki et al. 2002). 
  The various ingredients used  in this modeling were not necessary accurately
known, most noticeably the temperature mass ($T-M$) relation,
and the luminosity-temperature ($L-T$) relation and its possible  evolution with redshift. 
Therefore doubts have been raised up on  the 
applicability of this test  given these uncertainties 
 (Colafrancesco et al. 1997; Rosati et al. 2002).

The XMM-$\Omega$ project was conducted in order to provide an accurate 
estimation of the possible evolution of the luminosity temperature relation at
high redshift  for clusters of medium luminosity
which constitute the bulk of X-ray selected samples, allowing to remove a
major source of degeneracy in the determination of $\Omega_M$
(Bartlett et al. 2001). To maintain a better control on the resulting $L-T$ 
it is clearly preferable 
to use clusters 
homogeneously obtained from  X-ray selected samples. Therefore,
the choice of the SHARC surveys in the XMM-$\Omega$ project provides several advantages:  large angular coverage with  the Bright SHARC (Nichol et al. 1999; Romer et al. 2000),  deepness of the sample with the South SHARC (Burke et al. 
1997), still keeping the number of clusters  to a  realistic size for comprehensive 
X-ray investigations.

The purpose of this letter is to examine the expected number counts 
in  
comparison with the observed counts in two different cosmological models, namely
a concordance model and a high matter density flat universe,   and to examine the amplitude 
of the major sources of uncertainties: the statistical dispersion on 
the value of $\sigma_8$ from the finite local sample, the systematic 
uncertainties in the $M - T$ and the
$L - T$ relations at high redshift. In this modeling we take advantage of
the accurate knowledge of the $L-T$ relation provided by the
 first results from the XMM-$\Omega$ project 
(Lumb et al. 2003, hereafter L03) and from recent Chandra measurements of 
distant clusters (Vikhlinin et al. 2002).

\begin{table}
\begin{tabular}{lllllc} \hline
 
 T$_{15}$   & $\Omega_M$ & $\sigma_8$ & $\Gamma$ & Cosmological model  \\ 
 (keV)      &            &            &          &    and ingredients \\ \hline
   4         &  1.      &    0.55   &   0.12     &  A: best model+BN98+SMT\\
 6.5      &  0.85       &    0.455   &   0.1     &  A: best model+M98+SMT\\ 
 4         &  0.3       &    1.  &   0.2      & B: Low $\Omega_M$+BN98+SMT \\ 
 6.5      &  0.3       &    0.725   &   0.2     &  B: Low $\Omega_M$+M98+SMT\\ \hline
\end{tabular}
\caption{Models and parameters used in the number counts calculations.
}
\label{tab2}
\end{table}                       

\section{Modeling the temperature distribution function and number counts}

 Although a full 
likelihood is possible in order to determine the best parameters and 
their final uncertainty taking into account all the possible sources of 
uncertainties, we find enlightening to illustrate the differences in 
the counts predicted in the  two following specific models: the first model 
(hereafter model A) is the best flat model obtained as in B00 by fitting 
the local TDF and the high redshift TDF from the Henry (1997) sample, 
assuming a $\Gamma$ CDM--like 
spectrum. The second model (hereafter model B) we use is the so-called 
concordance model i.e. a flat $\Lambda$ CDM model with $\Omega_M = 0.3$, in 
agreement with the recent WMAP results (Spergel et al. 2003). In this model, 
only the local TDF was then fitted.
The parameters used are summarized in Table 1.
The samples we used for comparison are  EMSS (Gioia et al. 1990; Henry et al. 
1992), RDCS (Rosati et al. 1995, 1998), 160deg$^{2}$ 
(Vikhlinin et al. 1998, 2002), Bright SHARC (Nichol et al. 1999; Romer et al. 2000) and  MACS (Ebeling 
et al. 2001) for which the  selection criteria are believed to be well
known, generally given as sky coverage versus flux limit.
These catalogs provide us with a  sample of 274 (non necessarily independent) 
clusters with redshift between 0.3 and up to more than 1 and luminosities between 
$10^{43}$ and $10^{45}$ erg/s. 

In our modeling we follow a procedure close to the one used by Oukbir
\& Blanchard (1997), Reichart et al. (1999) and Borgani et al. (1999).
In the first step, models are normalized using the local TDF, 
which request two fundamental ingredients: the
mass function and the $M-T$ relation, assumed to follow a standard
scaling law (Kaiser 1986).
Intensive numerical simulations
have allowed to provide accurate analytical fits to the mass function (Sheth, 
Mo \& Tormen 2001, hereafter SMT; Jenkins et al. 2001; White 2002). Here we use the expression from SMT:
\begin{equation}
\frac{dN}{dm}\,=\,\sqrt{2a\over \pi}\,c\,\frac{\overline{\rho}}{m}\frac{d\nu}{dm}\,\left(1+{1\over (a\nu^2)^p}\right)
\,\exp\left(-{a\nu^2\over 2}\right)\,
\end{equation}
 with $a = 0.707$, $c = 0.3222$ and $p = 0.3$ and $\nu=\frac{\delta}{\sigma(m)}$.
 
\subsection{The $M - T$ relation} 
 It has been shown that the normalization of the $M-T$ relation based on 
numerical 
simulations is significantly different from the normalization inferred from 
hydrostatic equation (Roussel et al. 2000). In this work we use two 
different normalizations so as to cover the whole range of possibilities: 
 we use on one side the calibration based on the numerical simulations of 
Bryan and Norman
(1998, BN98 hereafter) and on the other side we use a $M-T$ relation 
 derived from the hydrostatic equation (Markevitch  1998, M98 hereafter) 
which produces a lower normalization of the matter power spectrum $\sigma_8$ 
(Seljak 2002; Reiprich \& B\"ohringer 2002). 
The $M-T$ relation is written to be:
\begin{equation}
T = T_{15}(\Omega_M \Delta(z,\Omega_M)/178)^{1/3}M_{15}^{2/3}(1+z)
\end{equation}
where $\Delta(z,\Omega_M)$ is the contrast density with respect to the density 
of 
the Universe for virialized objects (hereafter $h=0.5$). The subscript 15 means 
that  masses are taken 
  in unit of $10^{15}$ solar masses. BN98 found  $T_{15} \sim 3.8$ keV, while M98 
concluded to  a higher normalization : $T_{15} \sim 6.5$ keV.

 The mass
function can then be normalized from the observed TDF 
(a possible dispersion of 15\% is included as done in B00).
The availability of bright ROSAT clusters samples 
has allowed reasonably accurate estimations of the local TDF  (M98; 
 B00; Pierpaoli et al. 2001;
Ikebe et al. 2002), even if the agreement is partly due to 
the fact that the samples used in these previous works contain  nearly the same clusters. 

In the present work,
we use the local TDF based on an updated version of the local X-ray clusters 
sample used in B00 in which clusters were selected with fluxes 
$f_x > 2.2 \times 10^{-11}$ erg/s/cm$^2$ in the ROSAT
[0.1-2.4] keV and galactic coordinates $|b_{\rm II}|> 20^{\rm o}$. 
This revised TDF is essentially identical to B00 and other recent measurements.

\subsection{Evolution  of the $L-T$ relation}

\begin{figure}[t]
\begin{center}
\includegraphics[angle=0,totalheight=6.98cm,width=8.5cm]{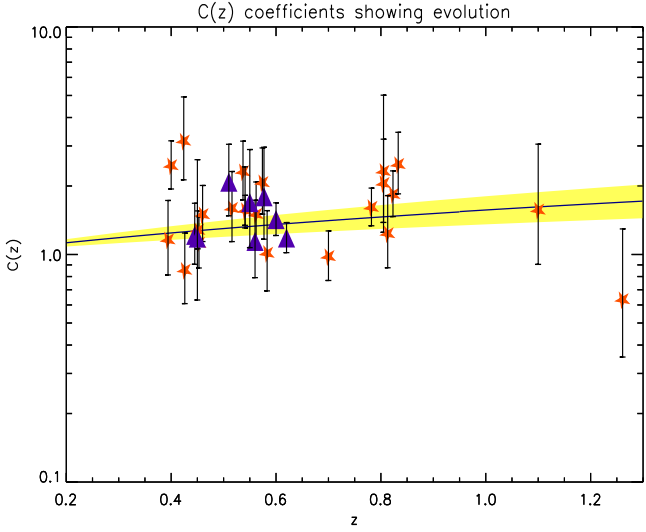}
\end{center}
\caption{\label{fig_powplot}  XMM measurements of the 
evolution of clusters $L-T$ relation expressed by the $C(z)$ coefficient (Eq. \ref{c(zi)}). Triangles are our XMM data and error bars are derived from the 1$\sigma$ error on the temperature measurement. Grey (yellow) area represents the 1$\sigma$ error on the $C(z)$ fit and stars are the Chandra data.}
\end{figure}
                                                          
In a second step we use the $L-T$ relation and its evolution 
to compute the luminosity function at different  redshifts. 
Using the observed local $L-T$ relation, models fitting the observed local TDF
automatically provide a luminosity function matching the data (Oukbir, Bartlett \& Blanchard, 1997).
The local $L-T$ relation $L_{\rm bol} = A(T/1 {\rm keV})^B$
is estimated from the 
sample described above (Vauclair et al. 2003 in preparation).  We obtained $A = 0.0625 \times 10^{44}$ erg/s
assuming
 $B = 3.$  We then estimated the evolution of this relation using  our XMM data of high-$z$ clusters obtained in the XMM-$\Omega$ project (L03). Following SBO98 we compute for each cluster:
\begin{equation}
C(z) = \frac{L}{AT^B} \frac{D_l(\Omega_M=1,z)^2}{D_l(\Omega_M,z)^2}
\label{c(zi)}
\end{equation}
and fit  these $C(z)$ by a power-law evolution law:
\begin{equation}
  C(z) = (1+z)^\beta
\label{c(z)}
\end{equation}
 the best-fit parameter $\beta= 0.65 \pm 0.21$ was obtained by a standard 
$\chi^2$ fitting (See figure \ref{fig_powplot}) in good agreement with what 
was found previously (SBO98) and fully consistent with the Chandra results 
(Vikhlinin et al. 2002). 

\subsection{Number counts}

 In order to compute number counts, one can notice that 
the observations actually provide $z$ and $f_x$ (rather than the actual $L_x$ and $T_x$). For a flux limited sample
with a sky  coverage area $\Omega$ and a flux limit $f_x$    one has therefore to compute the following:
\begin{eqnarray}
N(>f_x,z,\Delta z) = & \Omega \int_{z-\Delta z}^{z+\Delta z} \frac{\partial N}{\partial z}(L_x> 4\pi D_l^2 f_x) dz \nonumber \\
= & \Omega \int_{z-\Delta z}^{z+\Delta z} N(>T(z))dV(z) \nonumber \\
= & \Omega \int_{z-\Delta z}^{z+\Delta z} \int_{M(z)}^{+\infty}
N(M,z)dM dV(z)
\end{eqnarray}
{\it where $T(z)$ is the temperature threshold corresponding to the flux $f_x$
as given by the observations, being therefore 
independent of the cosmological model}. 
  We then compute the theoretical number counts in redshift bins $\Delta z = 0.1$ for the various
clusters samples mentioned above accordingly to their respective selection function.
In order to compare the same cluster population at different redshifts, we have excluded from the faint ROSAT samples number counts at redshift 
smaller than $z = 0.3$ for which clusters have too low luminosity compared to those used to measure the $L-T$ evolution in this project. In addition at $z < 0.3$ clusters become significantly
 more extended than the detection cell in the EMSS sample, and are not 
considered as reliable in MACS, a further reason for 
restricting our comparison to higher redshifts.

\subsection{Comparison with observations}

\begin{figure*}[!ht]
\vspace*{3mm}
\begin{center}
\includegraphics[angle=0,totalheight=6.3cm,width=8.5cm]{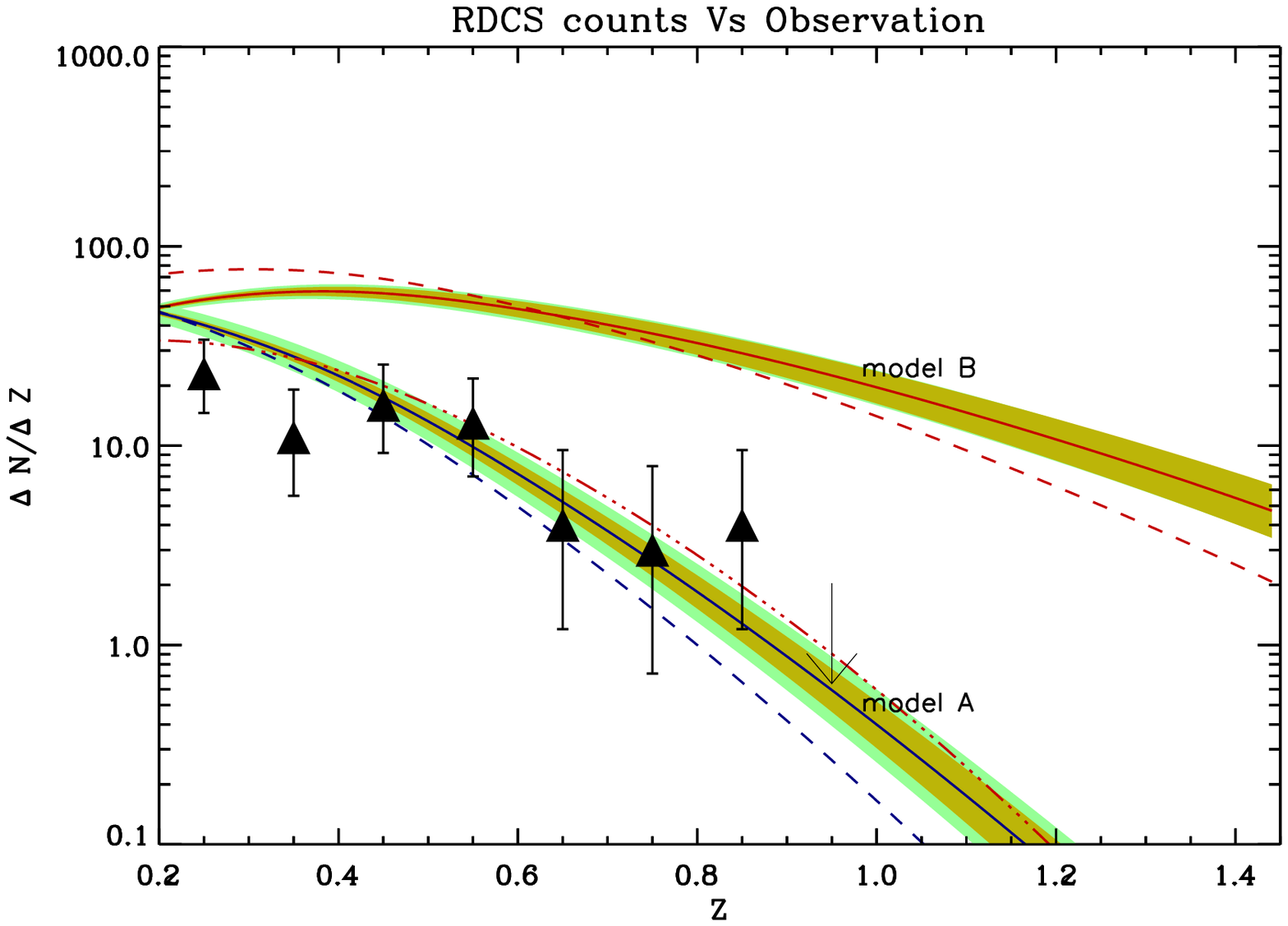}
\includegraphics[angle=0,totalheight=6.3cm,width=8.5cm]{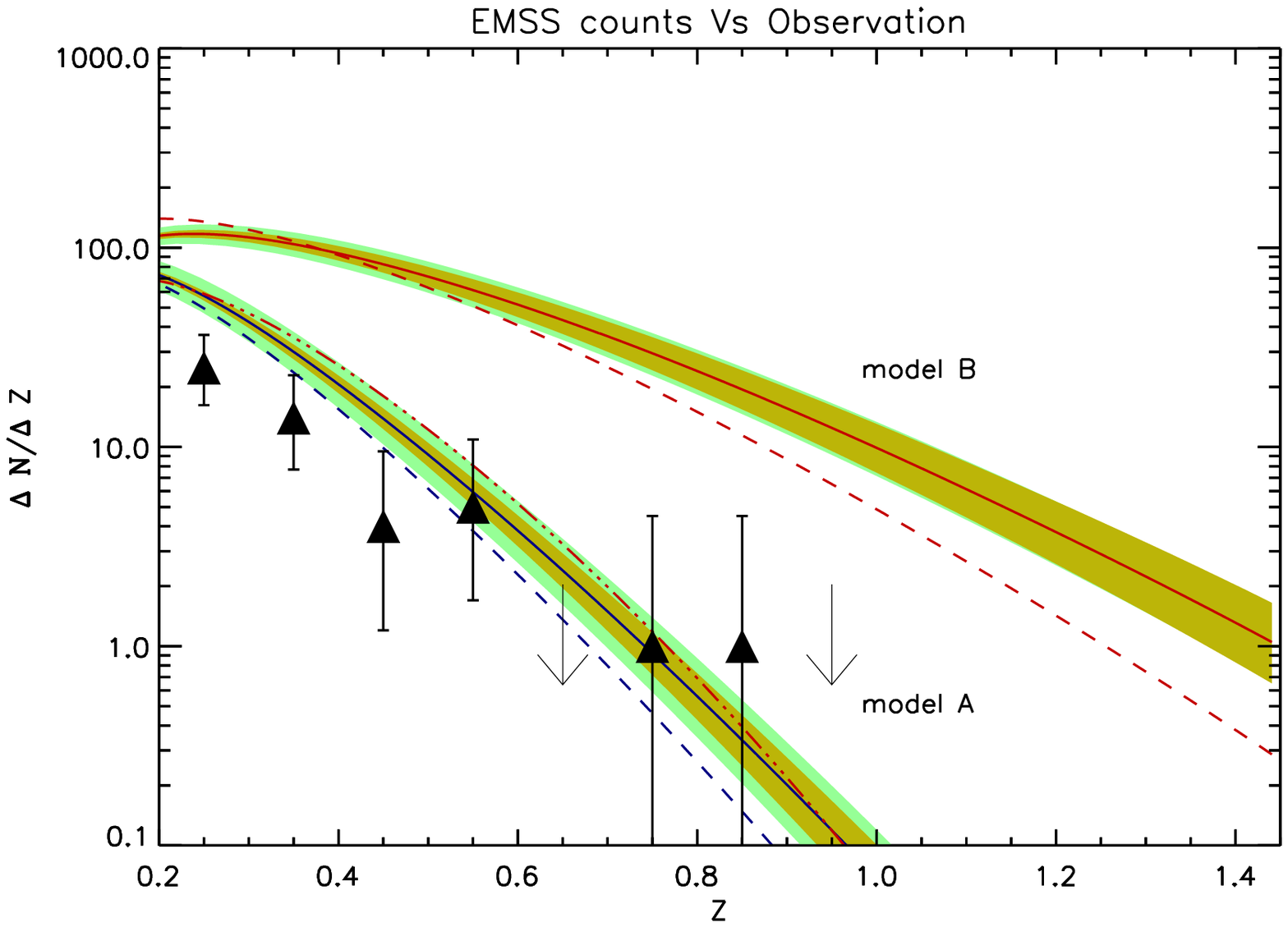} \\
\includegraphics[angle=0,totalheight=6.3cm,width=8.5cm]{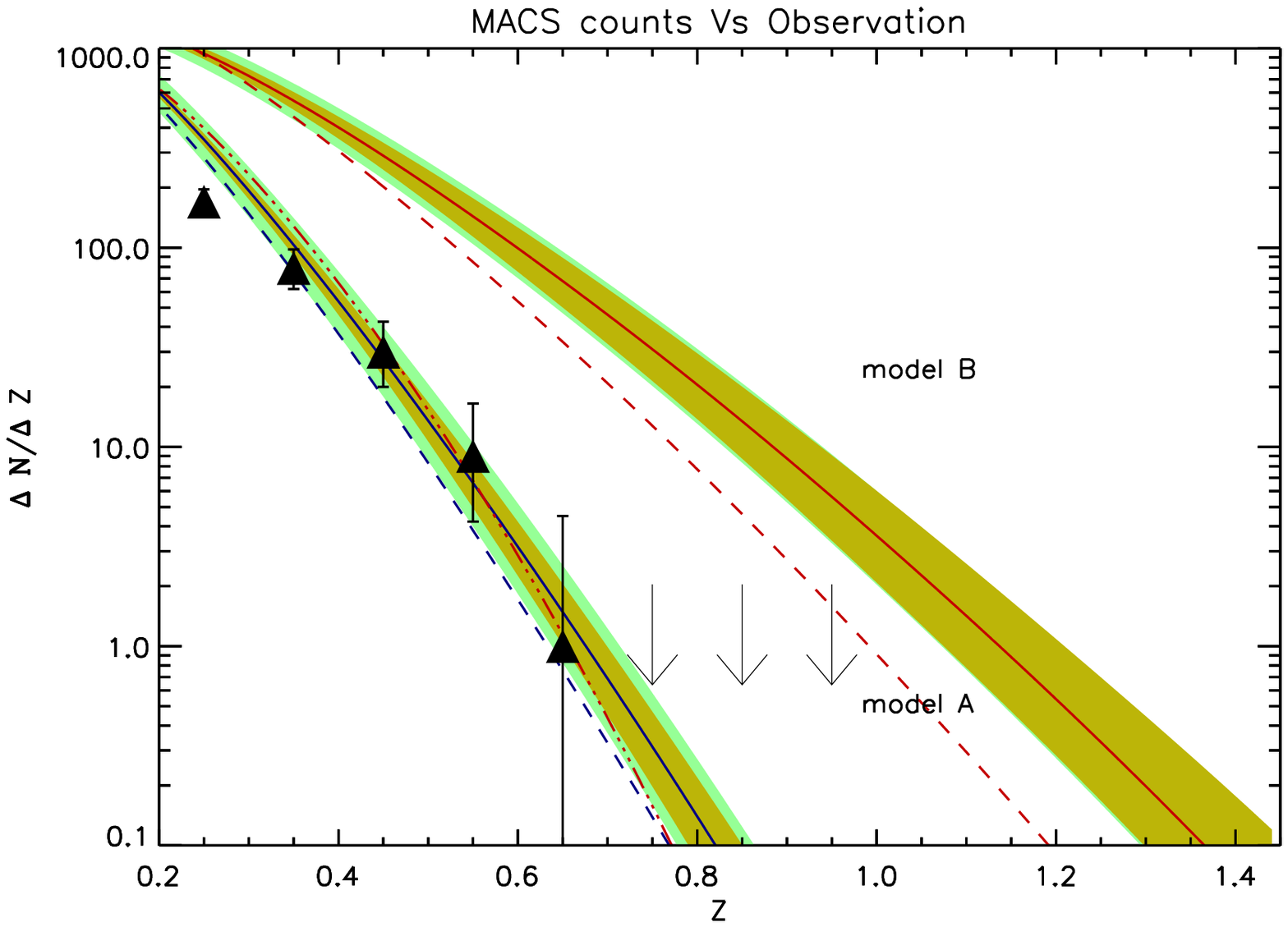}
\includegraphics[angle=0,totalheight=6.3cm,width=8.5cm]{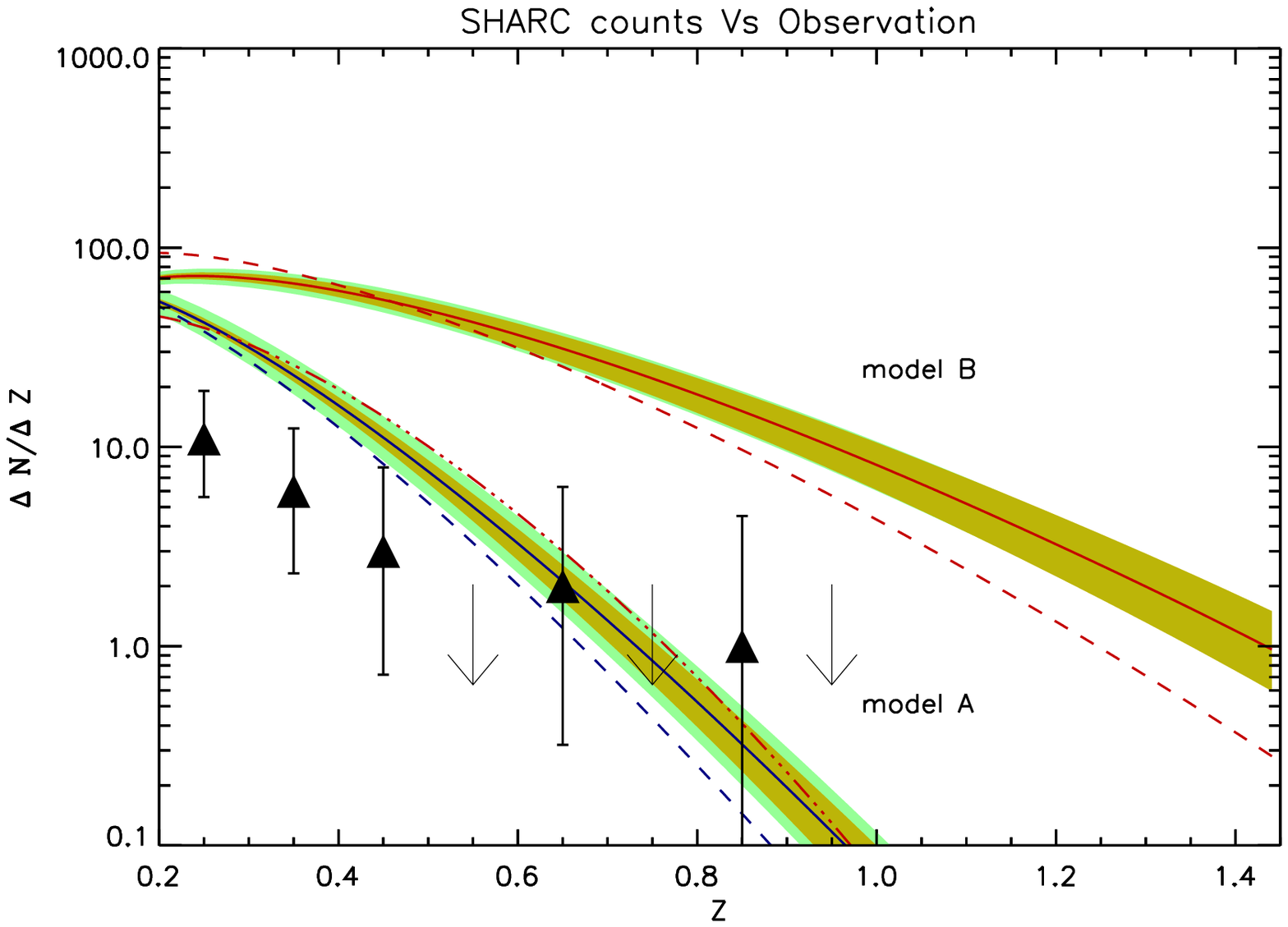}\\
\includegraphics[angle=0,totalheight=6.3cm,width=8.5cm]{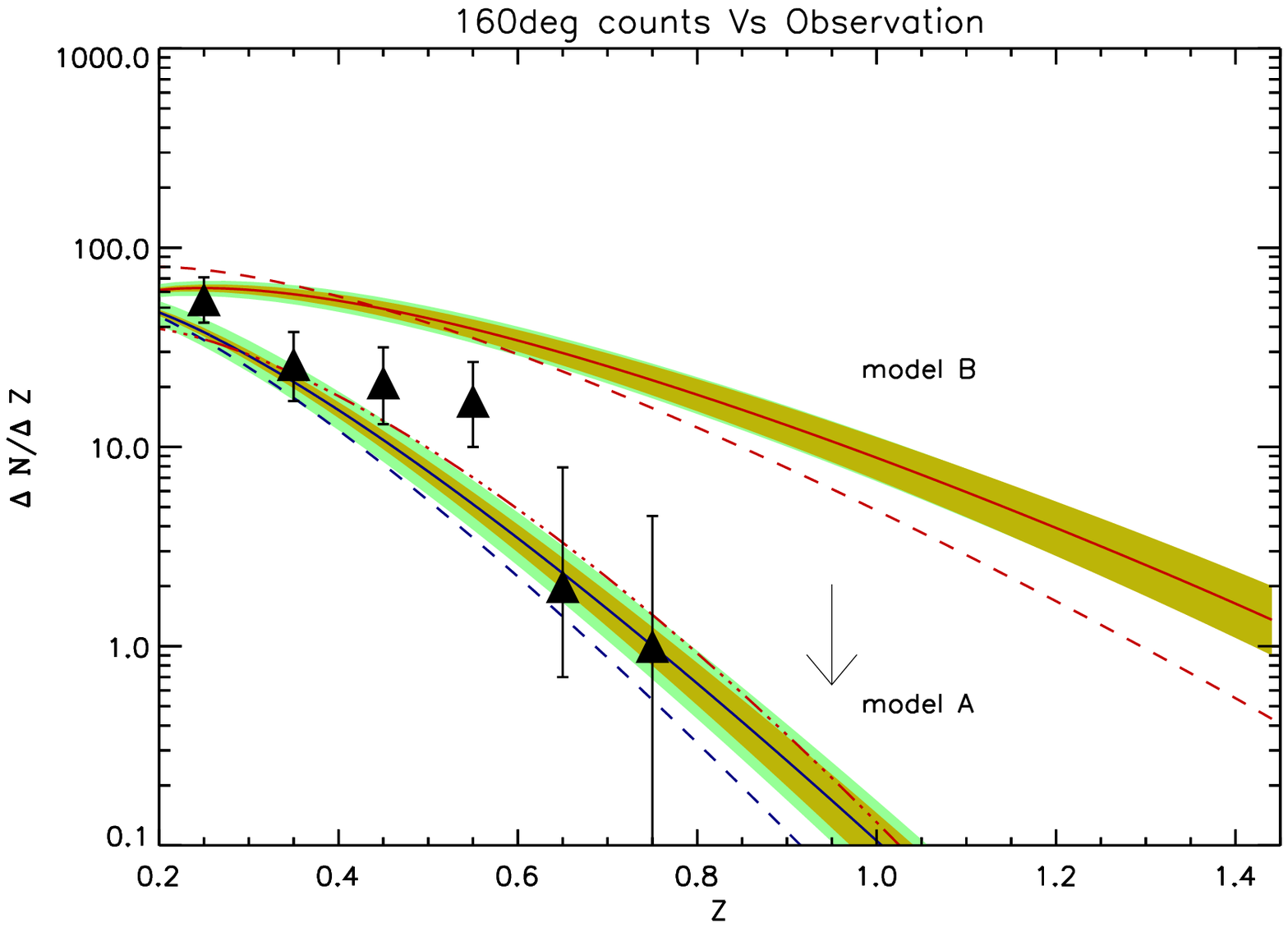}
\includegraphics[angle=0,totalheight=6.3cm,width=8.5cm]{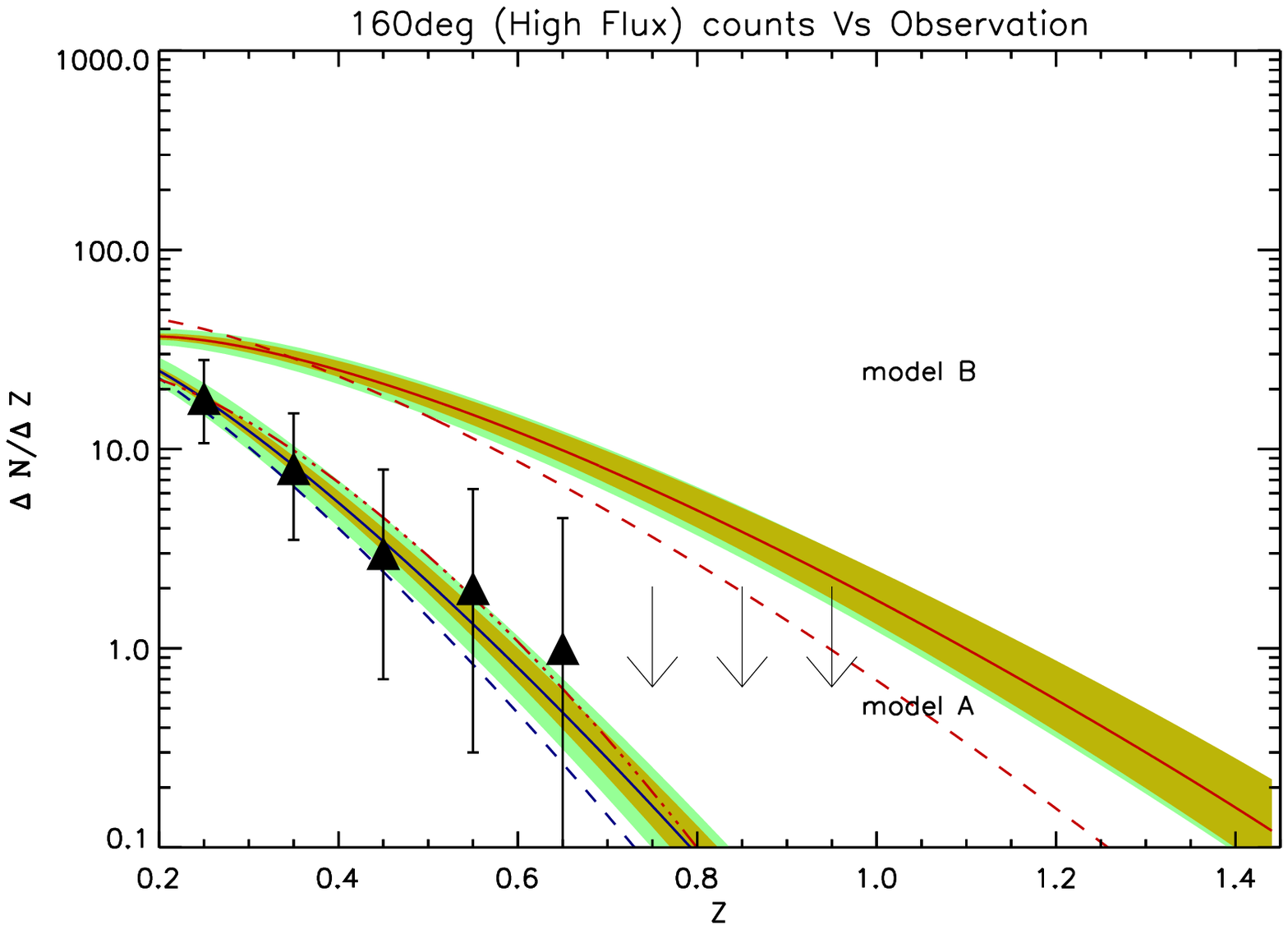}
\vspace*{3mm}
\caption{\label{xmm1} Theoretical  number counts in bins of redshift ($\Delta z=0.1$) for the  different surveys:
RDCS, EMSS, SHARC,  MACS and 160deg$^2$.  Observed numbers 
are triangles with 95\% confidence 
interval on the density assuming poissonian statistics (arrows are 95\% upper limits). 
 For the  160deg$^2$ survey we have  
also examined  the counts for the brightest part (160deg$^2$(high flux) corresponding to fluxes  $f_x > 2 \times 10^{-13}$ erg/s/cm$^2$), 
in order to show that the excess seen at $z\sim 0.5$ is not due to luminous clusters. The 
upper (red) curves are the predictions in the concordance 
model (model B). The lower (blue) curves are for model A (see table 1.). 
Different $M-T$ are figured: the dashed lines correspond to 
$T_{15} = 6.5$keV (M98 $M-T$)  while the continuous lines are for $T_{15} = 4.$keV,
corresponding to virial mass nearly twice larger, close to BN98 normalization.
 The dark  grey (green) area is the 
uncertainty range from our estimates on the uncertainty in the evolution 
of the $L-T$ relation while the light grey area is the 
uncertainty  on number counts due to  the dispersion on $\sigma_8$. The 
3-dotted-dashed lines show the predicted counts in the concordance model using 
$M-T$ relation from Eq. 6 violating the standard scaling with redshift. 
}
\end{center}
\end{figure*}

The final number counts 
for the different samples are presented in figure \ref{xmm1} for models A and B. As one can see the predicted counts 
in model A are 
in remarkable agreement with the observed number counts given the fact that no adjustment 
was performed, while the predicted
number counts in the concordance model reveals a strong disagreement 
with the data.  The  overproduction of clusters at high redshift is impressively large,
reaching a factor of nearly ten at redshift greater than 0.5.
Two factors are responsible for this difference:
the primary effect comes from the evolution in the number of clusters 
which is quite different in
the two models and the second effect comes from the increase of the volume element in the 
$\Lambda $ dominated universe.
The uncertainties in the value of $ \sigma_8$, in  the 
$L-T$ relation and in the $M-T$ relation  represent similar 
 uncertainties in the predicted counts, in the range of 30\% to 70 \%,
but are far below 
the difference between the two models which is between 5 and 10. 
We also checked that changing the slope and the normalization of the local $L - T$ does not affect our results.  
  One may worry whether the selection procedure of the samples
has introduced more bias than usually assumed. A systematic 
bias in the flux of the order of 2 to 3 for the different surveys would 
eliminate most of the discrepancy between the predictions 
of concordance model and the observed counts. Such a possibility seems 
very unlikely: different flux inter-comparisons exist, including those
of L03 between ROSAT SHARC and XMM fluxes and those of Vikhlinin et al. 
(2003) between ROSAT  and Chandra fluxes, and do not provide 
 any evidence for such a large bias.

\section{Discussion}

As we have seen a model which is normalized to the 
 local as well as to the high redshift TDF, reproduces 
 impressively well the redshift distribution of all the surveys 
we have investigated, without any adjustment and 
with little  uncertainties arising from the 
 modeling. In contrast, we have checked that relaxing the  
request of matching the local abundance  a concordance model fitting the redshift distribution leads
to a local  abundance of  X-ray clusters  $N(>4 {\rm keV},z = 0.05)$  nearly 
5 times lower than 
actually observed.
This is a  strong indication that existing samples of 
clusters (namely the Henry sample, the RDCS, the
EMSS, the Bright SHARC, the 160deg$^2$ and the MACS sample) draw the same 
picture, consistently pointing out toward the fact that the cluster abundance is significantly evolving with redshift, perhaps closing a long
term controversy on this question (Henry et al. 1992). Furthermore,
in standard hierarchical picture of structure formation 
such evolution points toward a high matter density universe with
$\Omega_M$ in the  range [0.85-1.], the precise value  depending on the 
$M-T$ normalization.
This conclusion is clearly conflicting with the currently popular 
concordance model. However, it should be emphasized that this  
is entirely consistent with {\em all previous existing analyzes}
performed on the redshift distribution of  X-ray selected samples 
of  clusters performed with the same methodology:   
our conclusion corroborates previous analyzes
of the EMSS clusters redshift distribution:
SBO98 found $\Omega_M=0.85\pm{.2}$, Reichart et al. (1999) found $\Omega_M=0.96\pm{.3}$  as well as the RDCS redshift 
distribution for models normalized to the present-day X-ray clusters abundance.
Indeed, from Eq. (14) in Borgani et al. (1999), $\Omega_M=(\beta+2)/3\pm{1/3}$
 we derive $\Omega_M \sim 0.88\pm{.34} $  from our $\beta \sim 0.65\pm{.21}$.
Note that high $\Omega_M$ models are also consistent with WMAP (Blanchard et al. 2003). 
A possible loophole could be 
 a large systematic bias flux, 
but we have argued that this seems very unlikely.
An other possibility would be that the scaling in the redshift of the $M-T$
relation (Eq. 1) is completely wrong, violating the basic scaling
scheme. Voit (2000)
 has investigated such 
a possibility, but concluded to a  moderate effect.
In  Fig. 2, we have plotted the predicted counts in a concordance model,
assuming 
\begin{equation}
 T = T_{15}(\Omega_M \Delta(z,\Omega_M)/178)^{1/3}M_{15}^{2/3}
\end{equation}
 instead of Eq. 2. As one can see, such a modification 
reestablishes agreement of the concordance model with observations. It is well known that 
 the $L-T$ relation cannot  be explained from simple scaling arguments.
One may therefore argue that the redshift evolution of the $M-T$
relation may suffer from more  dramatic effect than usually assumed, although -- to our knowledge -- 
such a possibility has never been advocated and it is probably 
not obvious to find
physical motivation leading to gas thermal energy in distant 
clusters
($z\sim 1$) to be  reduced by a factor of two compared to  clusters
in the local universe. 
 We therefore conclude that 
 the redshift distributions of present--day available 
X-ray clusters
surveys, as well as the recent results on the $L-T$ relation of high 
redshift clusters, favor a high matter density universe unless the standard 
paradigm on clusters gas physics has to be deeply revised. 

\begin{acknowledgements}
The authors would like to thank M.~Arnaud, J.~Peebles and  K.~Romer for 
fruitful comments during this work, H.~Ebeling and S.~Borgani for providing
us number counts of MACS and RDCS respectively. 
This research has made use of the Clusters Database BAX which is operated by the LAOMP under contract with the Centre National d'Etudes Spatiales (CNES).
\end{acknowledgements}

\end{document}